# The play of colours of prisms

**Amelia Carolina Sparavigna**
Institute of Fundamental Physics and Nanotechnology
Department of Applied Science and Technology
Politecnico di Torino, C.so Duca degli Abruzzi 24, Torino, Italy

*A short history of prisms from Lucius Anneus Seneca to George Ravenscroft*

The first scientific paper written by Newton was on the compound nature of white light, a paper where he described his experiments with prisms and the spectrum produced by the light. Newton had performed the experiments, probably between 1664 and 1667, in Woolsthorpe and Cambridge, using several glass prisms [1]. These glass artifacts existed and were sold at the country fairs to be used for the "play of colours", that is, to see the rainbow colours looking through them. This phenomenon of white light dispersion, created by crystals and glasses, was well known from ancient times and reported by Latin writers. Newton, experimenting with glass to improve optical instruments, explained that the play of colours was inside the nature of light. Few years after the Newton's studies, the development of lead glasses allowed new applications for the play of colours. Here a short history of prisms from Seneca to Ravenscroft.

## 1. Seneca, Pliny and their prisms
A Latin writer, Lucius Anneus Seneca, was fascinated by prisms. In his Natural Questions, [2,3], he told that there existed some glass rods commonly made in two forms: striated, and with many angles. If they received the sun's rays obliquely, they create the colours as it is usually seen in a rainbow. It seems therefore that the glassmakers were producing such rods for observing the play of colours.
About the phenomenon of the light dispersion, we have also the words written by Pliny in his Natural History. Pliny discusses of a precious stone, that he knew as Iris. He tells that the Iris is found in a certain island of the Red Sea, forty miles distant from the city of Berenice. He continues telling that the stone takes its name, Iris, from the property which it possesses; "for, when struck by the rays of the sun in a covered spot, it projects upon the nearest walls the form and diversified colours of the rainbow; continually changing its tints, and exciting admiration by the great variety of colours which it presents... The stone, however, as already stated, only presents these colours when under cover; not as though they were in the body of the stone itself, but, to all appearance, as if they were the result of the reflected light upon the surface of the wall. The best kind is the one that produces the largest arcs, with the closest resemblance to the rainbow." [4]
According to John Bostock that translated from Latin, the Iris, which means "rainbow", can be a prismatic crystal of limpid quartz. It is therefore possible that Roman glassmakers created glass rods having many angles to mimic the quartz crystals, which were quite precious objects.
It is interesting the "experimental set-up" of a covered spot to project the light on a nearest wall that Pliny describes.

## 2. Newton and the spectrum
It seems that Isaac Newton found some glass prisms sold at the country fairs, to conduct his first optical experiments to study the white light. We know this and many other details about these experiments from the book written by Brewster.

At the time Newton started his studies (1664-1667), as Brewster is telling, it had been long known that lenses with spherical surfaces did not give distinct images of objects [5]. "This indistinctness was believed to arise solely from their spherical figure... and various methods were suggested for diminishing or removing this source of imperfection. Descartes had shown that hyperbolic lenses refracted the rays of light to a single focus, and we accordingly find the early volumes of the Philosophical Transactions filled with schemes for grinding and polishing lenses of this form. Newton had made the same attempt, but finding that a change of form produced a very little change in the indistinctness of the image, he thought that the defect of lenses, and the consequent imperfection of telescopes, might arise from some other cause than the imperfect convergency of the incident rays to a single point. This happy conjecture was speedily confirmed by the brilliant discovery of the different refrangibility of the rays of light .... After our author (Isaac Newton) had purchased his glass prism at Stourbridge Fair, he made use of it in the following manner. Having made a hole H in his window-shutter SHT, and darkened the room, he admitted a ray of the sun's light RR, which after refraction at the two surfaces AC, BC of the prism ABC, exhibited on the opposite wall MN what is called the Solar or Prismatic Spectrum. This spectrum was an elongated image of the sun about five times as long as it was broad, and consisted of seven different colours, Red, Orange, Yellow, Green, Blue, Indigo, and Violet."

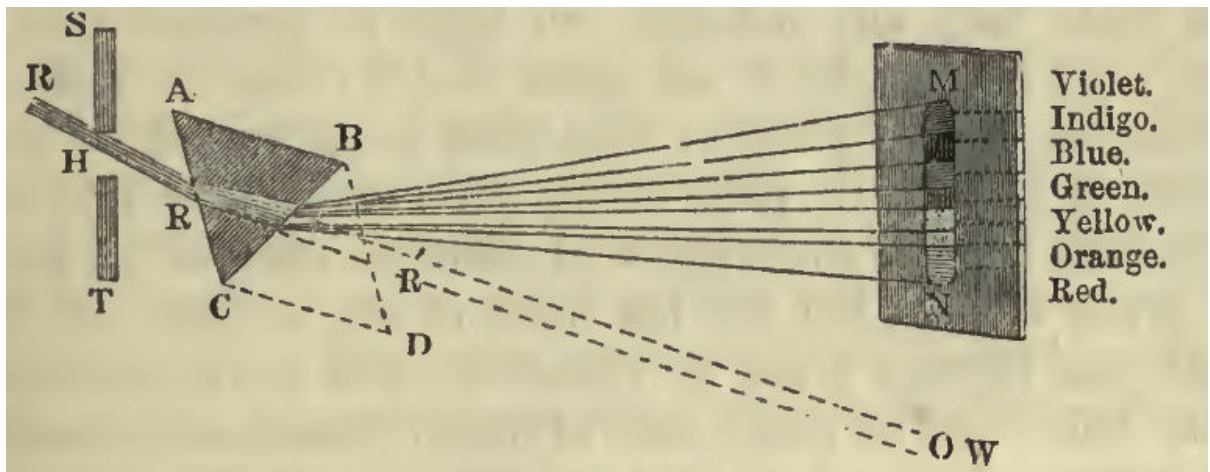
Illustration from the book by David Brewster

Then he took another prism BCD, and he observed that the light which by the first prism was diffused into MN, was reduced by the second prism into a circular OW. After this fact, Newton argued that whatever was the cause of image MN, it did not arise from any irregularity in the prism. He concluded that light was not homogeneous, but consisted of rays of different "refrangibility". "Having established this important truth, Newton immediately perceived that the different refrangibility of the rays of light was the real cause of the imperfection of refracting telescopes... As soon as Sir Isaac saw this result of his discovery, he left off his glass-works, as he called his attempts to improve the refracting telescope, and, in the autumn of 1668, constructed a little reflecting telescope...".
Here we have a wonderful example of fundamental physics: we have not only the observation of the phenomenon, which we have already found in the Pliny's description, but an experiment on it, obtained by using the second prism.

## 3. Ravenscroft and chandeliers

In the medieval period, the crystals of quartz continued to be valued for their ability to play the colours of the rainbow [6]. In the 13th century, Roger Bacon obtained hexagonal crystals from Ireland and India, while Albertus Magnus got his samples from Germany and the region round the Red Sea. The Pliny's Iris, which was so precious for Romans, continued to be a precious stone for a long time. Also the Roman glasses for the play of colours continued to be produced assuming the shape of prisms, having the knobbed ends as those used by Newton [7].

Aiming to improve optical instruments, Newton experimented the light dispersion using some prisms of soda glasses [1]. Since their refraction index of 1.55 is remarkably high for soda glasses, this is interpreted as evidence of the use of lead oxides added as flux. As Ref.1 is telling, the leaded glasses were used well before the invention of the high-lead glass produced by Ravenscroft in 1676. George Ravenscroft was an English businessman in the glass import/export and production. He spent a long period of his life in Venice, at that time famous for glass artistic production, in particular for the Cristallo, a glass imitating the rock crystals, invented by Angelo Barovier in 1450 [8,9]. In Venice, Ravenscroft established a successful business with his brothers and probably learned some glassmaking techniques [10].

There is some debate over the Ravenscroft's invention on the use of lead in production of glasses [10]. Probably he applied all what he learned in Venice on glasses, experimenting that adding lead oxide gave the glass different features. May be, he found some suggestions in books, for instance, that written by Antonio Neri, a Florentine priest. Neri wrote a collection of books on the art of glass, L'Arte Vetraria; and one of them was about the lead glass. It was first published in 1612 and translated into English in 1662. It is therefore possible that this book was relevant for the production of lead crystal glass by George Ravenscroft. And Newton could have read this book too.

In any case, in 1674, Ravenscroft applied to King Charles II for a patent to be sole manufacturer of lead crystal glass in England. He produced his glass for a period of only five years, ending the business in 1679. He died within a few days after the expiration of his patent, and other glassmakers immediately began to manufacture the lead glass [11].

The addition of lead oxide to glass raised its refractive index and lowered its working temperature and viscosity. The brilliance of these crystals increased due to the high refractive index produced by the lead content. This glass had a dispersion which increased the play of colours of prisms. A proper cutting can exploit these features to create a brilliant effect.

As previously told, the use of lead glass was already known, but it was the English production that rendered the art of glass quite popular throughout Europe. The light scattering properties of this highly refractive glass were added to prismatic forms, leading to the creation of the crystal chandeliers. In fact, it was the first quarter of the 18th century that saw the chandeliers being adorned with prisms to increase the scattering of light by internal reflections and the play of colours on the surrounding walls. And "prism" is the term that is still used to define the glasses which are used for chandeliers.